\documentclass[useAMS,usenatbib]{mn2e}
\usepackage{psfig}

\newcommand{\super}[1]{$^{#1}$}
\newcommand{\sub}[1]{$_{#1}$}
\newcommand{\lya}{Ly$\alpha$}

\def\farcs{\hbox{$.^{\prime\prime}$}}

\title[Dust Emission from \lya~Galaxies]
{The Expected Detection of Dust Emission from High-Redshift Lyman Alpha Galaxies}

\author[Finkelstein et al.]
  {Steven   L.    Finkelstein$^{1,2}$\thanks{Email: stevenf@asu.edu}, 
  Sangeeta Malhotra, James E. Rhoads$^{3}$,  Nimish P. Hathi$^{4}$\newauthor and Norbert Pirzkal$^{5}$\\
  $^{1}$Department  of  Physics,  Arizona  State University,  Tempe, AZ  85287\\
  $^{2}$George P. and Cynthia W. Mitchell Institute for Fundamental Physics and Astronomy, Department of Physics, Texas A\&M\\
 University, College Station, TX 77843\\
  $^{3}$School of Earth and Space Exploration,  Arizona  State University,  Tempe, AZ  85287\\
  $^{4}$Department of Physics and Astronomy, University of California, Riverside, CA 92521\\
  $^{5}$Space Telescope Science Institute, Baltimore, MD 21218}

\date{Date Accepted.  Date received.}

\pagerange{\pageref{firstpage}--\pageref{lastpage}} \pubyear{2008}

\begin{document}

\label{firstpage}

\maketitle

\begin{abstract}
Recent results have shown that a substantial fraction of high-redshift Lyman alpha galaxies contain considerable amounts of dust. This implies that Lyman alpha galaxies are not primordial, as has been thought in the past.  However, this dust has not been directly detected in emission; rather it has been inferred based on extinction estimates from rest-frame ultraviolet (UV) and optical observations.  This can be tricky, as both dust and old stars redden galactic spectra at the wavelengths used to infer dust.  Measuring dust emission directly from these galaxies is thus a more accurate way to estimate the total dust mass, giving us real physical information on the stellar populations and interstellar medium (ISM) enrichment.  New generation instruments such as the Atacama Large Millimeter Array (ALMA) and Sub-Millimeter Array (SMA), should be able to detect dust emission from some of these galaxies in the sub-mm.  Using measurements of the UV spectral slopes, we derive far-infrared flux predictions for of a sample of 23 z $\geq$ 4 Lyman alpha galaxies.  We find that in only a few hours, we can detect dust emission from 39 $\pm$ 22\% of our Lyman alpha galaxies. Comparing these results to those found from a sample of 21 Lyman break galaxies (LBGs), we find that LBGs are on average 60\% more likely to be detected than \lya~galaxies, implying that they are more dusty, and thus indicating an evolutionary difference between these objects.  These observations will provide better constraints on dust in these galaxies than those derived from their UV and optical fluxes alone.  Undeniable proof of dust in these galaxies could explain the larger than expected \lya~equivalent widths seen in many \lya~galaxies today.
\end{abstract}


\section{Introduction}
Lyman alpha (\lya) galaxies have been studied in detail for nearly a decade now, and we are only just beginning to learn about the properties of individual objects.  Most initial narrow-band studies (e.g., Hu et al. 1998; Rhoads et al. 2000) observed blank fields with two filters: a narrow-band filter (to capture the \lya~emission line), and a broad-band filter encompassing this narrow-band (used to measure a narrow-band excess).  Learning the detailed properties of individual \lya~galaxies would be difficult with these data.  However, these early studies were able to measure the equivalent widths (EWs) of their objects, calculated from the ratio of narrow-band flux to broad-band flux (Malhotra \& Rhoads 2002).

Some of these initial results were surprising, as in many cases the \lya~EW distribution was higher than one might have expected from stellar population models, as many objects exhibited rest-frame \lya~EWs over 200 \AA~(e.g., Kudritzki et al. (2000); Malhotra \& Rhoads (2002); Dawson et al. (2004; 2007); Shimasaku et al. (2006); Finkelstein et al. (2007)).  A normal stellar population (defined as having a Salpeter (1955) initial mass function (IMF), solar metallicity, and a constant star formation rate) has a maximum \lya~EW of 260 \AA~at 1 Myr, falling to a constant value of 95 \AA~at 100 Myr.  Thus, while objects with EWs $\geq$ 200 \AA~are possible with a normal stellar population, they would all have to be extremely young.  Looking at a random sample of \lya~galaxies, it would be extraordinarily rare if they all had ages of only a few Myr, thus something is likely happening to skew the EWs higher than they should be.

There are many possible reasons why this could be occurring, and the easiest area to examine is the assumption of a normal stellar population.  For example, if a stellar population had sub-solar metallicity, it would form stars with hotter stellar photospheres.  These stars would then pour out more ionizing photons, which, after interacting with the local interstellar medium (ISM), would in turn create more \lya~photons.  Another possibility exists, whereby if a galaxy had a top-heavy initial mass function, it would create more massive stars than a Salpeter IMF, again resulting in more ionizing photons, and thus more \lya~photons.  Both of these scenarios are possible in primitive, dust-free galaxies.  Strong \lya~lines also come from the active galactic nuclei (AGN) components of some galaxies, but previous work has shown that most \lya~galaxies detected via narrow-band selection are not AGN (Malhotra et al. 2003; Wang et al. 2004).

\lya~photons are resonantly scattered by any neutral hydrogen present, thus in a homogeneous ISM, they will have to travel much further to escape the galaxy, dramatically increasing the chance of absorption by dust (Meier \& Terlevich 1981).  However, if the dust is in an inhomogeneous distribution, it is possible that a galaxy could appear to have a strong \lya~EW at later ages.  

In this scenario, \lya~can escape unaffected by dust if it is somehow screened from interacting with the dust grains.  This can happen if the dust resides in cold, neutral clumps embedded in an ionized inter-clump medium.  This scenario was first proposed by Neufeld (1991) and examined in detail more recently by Hansen \& Oh (2006).  In this ISM geometry, the \lya~photons get scattered by neutral hydrogen at the surface of these clumps, and thus never see the dust which is embedded deeper inside.  However, the continuum photons suffer no such restriction, so they pass un-obstructed into the interior of the clump, with many continuum photons being absorbed (extincted) and scattered (resulting in reddening of the continuum colour).  The \lya~photons proceed to bounce around from clump to clump, eventually escaping the galaxy.  The net result of this is an increase in the \lya~-- continuum flux ratio over that intrinsic to the stellar population.  We would thus detect this galaxy with a higher EW, irrespective of its stellar population characteristics.

We believe that this scenario may be responsible for some of the high-EW objects seen in narrow-band studies, and in our previous work (Finkelstein et al. 2009; hereafter F09), we found ten objects out of a sample of 14 LAEs where this scenario appears to be in effect (one of which was also studied in Finkelstein et al. 2008a).  Two of these objects are evolved galaxies with large \lya~EWs due to this dust enhancement scenario (the other eight are still young, but have their \lya~EWs enhanced to be larger than their already large value due to dust).  The first of these two evolved galaxies was best fit by a 450 Myr, 3.8 x 10\super{9} M$_{\odot}$ stellar population with a rest-frame EW of $\sim$ 100 \AA.  This relatively high EW at old age was due to dust, as the best-fit model had 0.3 magnitudes of dust extinction at 1200 \AA, but this dust was in a geometry such that it did not attenuate \lya~measurably.  This resulted in a $\sim$ 35\% increase in the \lya~EW over a similar stellar population with no dust.  The second of these objects had a most likely age (see F09 for derivation of most likely age) of 500 Myr, a mass of 5 $\times$ 10$^{10}$ M$_{\odot}$, and 3.5 mag of dust extinction in a very clumpy geometry, greatly enhancing the \lya~EW.

While these objects are very interesting, and could go a long way towards explaining the apparent discrepancy in EW distributions, the rest of the objects in the sample had their own surprises.  They were all young and lower-mass, with ages from 3 -- 15 Myr, and masses from 1.6 $\times$ 10$^{8}$ -- 6.1 $\times$ 10$^{9}$ M$_{\odot}$.  These ages and masses are indicative of the traditional belief of \lya~galaxies, where they are very young, and possibly primordial in nature.  However, all of these twelve LAEs were best fit by stellar populations with considerably non-primordial amounts of dust, with A\sub{1200} = 0.5 -- 4.50, with many galaxies showing indications of a clumpy ISM.

By fitting their observations to stellar population synthesis models, many other recent studies have derived dust extinction in their \lya~galaxies as well (e.g., Chary et al. 2005; Pirzkal et al. 2007; Lai et al. 2007; Pentericci et al. 2008).  However, these model parameters can be degenerate, as at rest-frame UV and optical wavelengths the reddening of the continuum colour could be due to both dust {\it and} old stars.  While observing these objects in the rest-frame optical with the {\it Spitzer Space Telescope} helps reduce this degeneracy, data at longer wavelengths could provide much tighter constraints.  

When UV and optical photons are absorbed by dust, they are re-emitted in the far-infrared (FIR), resulting in a second observed peak in the spectral energy distribution (SED).   Observing these galaxies in the sub-millimeter and millimeter (mm) regimes would detect this dust emission directly, which could resolve this red-colour degeneracy.  The Atacama Large Millimeter Array (ALMA) will be a uniquely sensitive tool to detect these high-redshift galaxies if they have dust, given its excellent observing conditions and large baseline.  Thus, obtaining ALMA observations (even non-detections), will provide a much tighter constraint on the amount of dust in these objects than rest-frame UV-optical observations alone.  We use ALMA as a measure of future observatory capabilities of detecting dust emission from LAEs.  If we can confirm that many \lya~galaxies do in fact have dust, then it will help show that they are not the primitive galaxies that they were once thought to be.  On the other hand, if we do not find evidence for dust, it may indicate that a top-heavy IMF or Population III stars are responsible for the strong \lya~EWs.

Where applicable, we assume Benchmark Model cosmology, where $\Omega_{m}$ = 0.3, $\Omega_{\Lambda}$ = 0.7 and H$_{0}$ = 0.7 (c.f. Spergel et al. 2007).  All magnitudes are listed in AB magnitudes (Oke \& Gunn 1983).  In Section 2 we discuss our sample of \lya~galaxies which we use to make predictions.  In Section 3 we present our results, and in Section 4 we make conclusions about the implications of these results.

\section{Method}

\subsection{\lya~Galaxy Sample}
We examine samples of LAEs from two recently published studies:  Finkelstein et al. 2009 and Pirzkal et al. 2007.  The sample from F09 consists of 14 \lya~galaxies in the Great Observatories Origins Deep Survey (GOODS) {\it Chandra} Deep Field -- South (CDF--S) narrow-band selected to be at z $\approx$ 4.5.  We will use the identifiers from F09, which are based on their detection filters:  CHa-1 to 4 (detected in NB656, or $H\alpha$), CH8-1 and 2 (detected in NB665, or $H\alpha$ + 80), and CS2-1 to 8 (detected in NB673, or [SII]).  See F09 for further details on the data reduction and selection process.  Seven of the 14 objects have published photometric redshifts putting them near z $\sim$ 4 -- 4.5 (MUSIC, Grazian et al. 2006; FIREWORKS, Wuyts et al. 2008).  As mentioned in the introduction, 12 of these objects were best-fit by young (3 -- 15 Myr), low-mass (1.6 -- 60 $\times$ 10\super{8} M$_{\odot}$) stellar populations with a good amount of dust (A\sub{1200} = 0.5 -- 4.5), with the dust in some cases enhancing the \lya~EW.  Two objects (CHa-3 and CS2-4) were determined to contain older (450 -- 500 Myr), more evolved populations, with dust extinction attenuating the continuum only due to a clumpy ISM, thus enhancing the \lya~EW.

The sample from Pirzkal et al. (2007; hereafter P07) consists of nine \lya~galaxies selected in the {\it Hubble} Ultra-Deep Field (HUDF; Beckwith et al. 2006), identified by their emission lines with spectra from the GRAPES (Grism ACS Program for Extragalactic Science; Pirzkal et al. 2004; Malhotra et al. 2005; Xu et al. 2007; Rhoads et al. 2008) survey.  These objects were selected based on a detectable \lya~emission line in the GRAPES grism data, and broadband ACS photometry consistent with a Lyman break galaxy (P07).  In this paper, we will use the object IDs from P07: 631, 712, 4442, 5183, 5225, 6139, 9040, 9340 and 9487.  These objects lie in the redshift range 4.00 $\leq$ z $\leq$ 5.76, with \lya~line fluxes ranging from 17 -- 60 $\times$ 10\super{-18} erg s\super{-1} cm\super{-2} and $z'$ magnitudes from 25.66 -- 28.36.  Using an exponentially decaying star formation history (SFH), P07 found best-fit ages of 1 -- 20 Myr and stellar masses from 7 $\times$ 10\super{6} -- 1.8 $\times$ 10\super{9} M$_{\odot}$.  Three of the objects were best fit by dust, with 0.05 -- 0.6 magnitudes of extinction in the V-band (four objects were best-fit with dust using a single-burst SFH, and five objects were best-fit with dust using a two-burst SFH).  

For comparison with our LAEs, we also use the catalog of spectroscopically confirmed HUDF Lyman Break Galaxies (LBGs) from Hathi, Malhotra \& Rhoads (2008; hereafter HMR08; Malhotra et al. 2005; Rhoads et al. 2008).  There are 49 LBGs in the catalog, but for objects at redshifts greater than 4.5, we require J and H band measurements to derive their FIR fluxes, thus we were restricted to using only 21 of the 49 LBGs.

\subsection{Dust Emission}
While we know that photons absorbed by dust are re-emitted in the far-infrared, the stellar population models used in both F08ab and P07 (Bruzual \& Charlot 2003), while computing fluxes as red as 160 $\mu$m rest-frame,  do not compute dust emission.  In order to predict the FIR dust emission from the objects in our sample, we exploit a relation between the UV spectral slope ($\beta$; defined as f\sub{\lambda} $\propto$ $\lambda^{\beta}$ by Calzetti et al. 1994) and the infrared excess (defined as the ratio of FIR flux to UV flux) detailed in Meurer et al. (1997; hereafter M97).  This relation is shown in Figure 1 of M97, with the infrared excess increasing exponentially with the spectral slope.  This relation is due to dust reddening of the spectral slope ($\beta$ becoming less negative).  Larger quantities of dust will increase the amount of UV extinction, increasing the amount of flux re-emitted in the FIR.  

Since we have obtained estimates of dust in our previous studies, we could use our derived extinction values to estimate the FIR flux.  However, not all objects are able to be detected in the observed IR, especially at higher redshift.  These IR data are necessary to break the aforementioned degeneracy of red colors due to dust or old stars.  We aim to predict FIR fluxes independent of stellar population modeling results, with the goal that our method can be applied to objects which only have detections in the optical or NIR.

To use the M97 relation, we first require the UV spectral slope for all objects in our sample, which usually requires spectra.  However, in M97, the authors detail a method of obtaining $\beta$ from the broad-band V and R magnitudes.  Unfortunately, this relation is only valid for z $\sim$ 3, but HMR08 compute similar relations for z $\sim$ 4 and z $\sim$ 5 -- 6.  For objects in our sample with z $\leq$ 4.5, we computed $\beta$ via:
\begin{equation}
\beta = 5.65 \times (i' - z') - 2
\end{equation}
and for objects with z $>$ 4.5, we computed $\beta$ via:
\begin{equation}
\beta = 2.56 \times (J - H) - 2
\end{equation}
Thirteen of the 23 objects in our sample were calculated to have $\beta$ $\geq$ -2.5, indicating that the dust in these galaxies is contributing significantly to their FIR fluxes.  Table 1 details the relevant parameters for each object in our sample, including the UV spectral slopes calculated with the above method.

Once $\beta$ is calculated, we use the relation from M97 to compute the FIR flux from the UV flux (defined in M97 as the flux at 2320 \AA).  First, we used the continuum slope calculated in equation 1 or 2, and then from the definition of continuum slope (see above), we calculated the UV flux via:
\begin{equation}
F_{UV} = f_{\lambda,z} \cdot \left(\frac{[2320 \cdot (1 + z)]^{\beta+1}}{9200^{\beta}}\right)
\end{equation}
where f$_{\lambda,z}$ is the z' flux density in units of f$_{\lambda}$ and 9200 \AA~is the central wavelength of the z' filter.  To find F\sub{FIR}/F\sub{UV}, we fit an exponential function to the dashed line in Figure 1 of M97.  The equation of this line was:
\begin{equation}
\frac{F_{FIR}}{F_{UV}} = 66.3 \cdot e^{1.41\cdot\beta} - 1.92
\end{equation}
This equation accurately matched M97's Fig. 1 over the range of -2.5 $<$ $\beta$ $<$ 4.  While some objects in our sample did have $\beta$ $<$ -2.5, these objects will have a very small F$_{FIR}$/F$_{UV}$ ratio, and thus the emission from their small amount of dust will very likely not be detectable.  

In order to obtain the FIR flux densities at a few representative wavelengths, we used equation 10 from M97, which details how the FIR flux is composed of the flux densities at 60 $\mu$m and 100 $\mu$m (in Jy):
\begin{equation}
F_{FIR} = 1.26 \times 10^{-11} (2.58f_{60} + f_{100})~\mbox{erg~cm$^{-2}$~s$^{-1}$}
\end{equation}

To independently calculate f\sub{60} and f\sub{100}, we require second equation relating these two parameters.  We used a template SED from Bendo et al. (2006), who computed the best fit modified blackbody curve to NGC 4631, an edge-on Sd galaxy (a reasonable local analog to LAEs).  They found that a simple blackbody modified by a $\lambda^{-1.2}$ term, with a dust temperature of 28 K, fit their observed data points from $\sim$ 10 -- 1000 $\mu$m.  
\begin{equation}
B_{\nu}(T) \propto \frac{2h\nu^{3}}{c^{2}}\frac{1}{e^{h\nu/kT} - 1}\left(\frac{c}{\nu}\right)^{-1.2}
\end{equation}
Adopting an SED of this form and T = 28 K, we calculated a ratio of f$_{100}$/f$_{60}$ of 3.64.  Using this in combination with Eqn. 5, we found:
\begin{equation}
f_{60} = \frac{F_{FIR}}{1.26 \times 10^{-11}(2.58 + 3.64)}~\mbox{Jy}
\end{equation}

\section{Results}
While we know the rest-frame 60 $\mu$m flux for the objects in our sample, we also would like to know how their FIR flux changes with wavelength.  To do this, we scaled the Bendo et al. modified blackbody function up to the value of the 60 $\mu$m flux for each object (redshifted to the redshift of the object), so that we could see how their fluxes behaved at all wavelengths in the range of interest.  We can then compare the flux densities of these objects at any FIR wavelength to the sensitivities of sub-mm instruments to see if an object will be detected.
\begin{table*}
  \centering
  \begin{minipage}{120mm}
    \caption[Properties of LAEs in the Sample]{Properties of LAEs in the Sample\footnote{This table lists some of the relevant properties of the LAEs in our sample. Objects with $\beta$ $<$ -2.5 were not able to have their FIR excess derived via the Meurer relation, and thus we do not tabulate their fluxes at 60 and 100 $\mu$m (with such a steep slope, they likely would not have detectable dust emission).}}\label{almatab}
    \begin{tabular}{@{\extracolsep{\fill}}ccccccc}
      \hline
      \hline
      \multicolumn{1}{c}{Name} & \multicolumn{1}{c}{Redshift} & \multicolumn{1}{c}{$\beta$} & \multicolumn{1}{c}{$f_{UV}$} & \multicolumn{1}{c}{$f_{60}$} & \multicolumn{1}{c}{$f_{100}$} & \multicolumn{1}{c}{Dust Detectable?}\\
      \multicolumn{1}{c}{$ $} & \multicolumn{1}{c}{$ $} & \multicolumn{1}{c}{$ $} & \multicolumn{1}{c}{(10$^{-20}$ erg cm$^{-2}$ s$^{-1}$)} & \multicolumn{1}{c}{($\mu$Jy)} & \multicolumn{1}{c}{($\mu$Jy)} & \multicolumn{1}{c}{(Y/N)}\\
      \hline
      CHa-1&4.40&-3.46&1.84&---&---&N\\
      CHa-2&4.40&-2.50&3.41&0.178&0.648&N\\
      CHa-3&4.40&-2.90&2.76&---&---&N\\
      CHa-4&4.40&-1.21&10.46&169.1&615.4&Y\\
      CH8-1&4.47&-0.35&3.97&248.1&903.2&Y\\
      CH8-2&4.47&-2.67&5.13&---&---&N\\
      CS2-1&4.53&-2.22&2.96&4.733&17.23&N\\
      CS2-2&4.53&-0.52&2.65&129.7&472.2&Y\\
      CS2-3&4.53&1.52&8.44&7784&28333&Y\\
      CS2-4&4.53&-1.14&0.65&12.05&43.87&Y\\
      CS2-5&4.53&-4.70&0.32&---&---&N\\
      CS2-6&4.53&-0.10&1.37&125.3&456.1&Y\\
      CS2-7&4.53&-1.65&0.44&3.296&12.00&N\\
      CS2-8&4.53&-2.66&0.70&---&---&N\\
      631&4.00&-2.34&2.28&1.779&6.477&N\\
      712&5.20&N/A&0.747&---&---&N\\
      4442&5.76&-4.94&0.145&---&---&N\\
      5183&4.78&-0.75&0.327&11.80&42.97&Y\\
      5225&5.42&-3.61&3.422&---&---&N\\
      6139&4.88&-2.03&4.191&13.63&49.61&Y\\
      9040&4.90&-1.87&3.093&15.27&55.58&Y\\
      9340&4.71&N/A&0.521&---&---&N\\
      9487&4.10&-2.62&0.822&---&---&N\\
      \hline
    \end{tabular}
  \end{minipage}
\end{table*}
With the ALMA sensitivity calculator\footnote[1]{Available online at: http://www.eso.org/projects/alma/science/bin/sensitivity.html}, we computed ALMA sensitivity curves using the default set of parameters for an integration time of four hours in bands 8 (385 -- 500 GHz) and 9 (602 -- 720 GHz).  Continuum measurements are possible with ALMA in 8 GHz instantaneous bandpasses, corresponding to a bandpass width of 14 $\mu$m at 730 $\mu$m (Wilson 2007).  We thus chose three windows in which to compare our objects to the ALMA sensitivity curves: 450, 610 and 730 $\mu$m.  The average sensitivity in an 8 GHz width across these three bandpasses are plotted as thick black lines in Figure 1.  From the relative positions of these sensitivities to the SED curves from our objects in this figure, we can see which objects will have their dust emission detected with ALMA.
%-----------------------Figure------1--------------------
\begin{figure}
\psfig{file=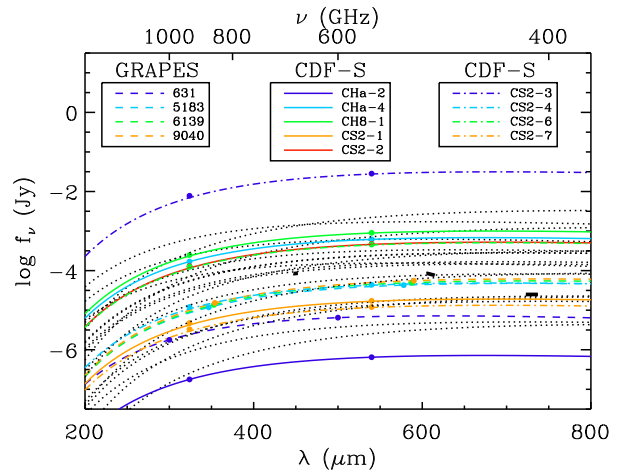,width=3.5in}
\caption{Derived FIR SEDs versus observed wavelength of our objects, with 13 LAEs being shown in the plotted range.  The different coloured lines show the LAEs, while the dotted black lines show the SEDs of the sample of LBGs.  The curves were derived by scaling the computed 60 $\mu$m flux point to a modified blackbody curve (Bendo et al. 2006).  The FIR fluxes of these objects were computed via the relation between the UV spectral slope and FIR excess from Meurer et al. (1997).  The thick black lines show the ALMA sensitivities in a four hour integration, in three windows at 450, 610 and 730 $\mu$m.  Out of the whole sample, nine LAEs (13 LBGs) would have their dust emission detected in a 4 hour ALMA integration.}
\end{figure}
%--------------------------------------------------------

\subsection{Lyman Alpha Galaxies}
Out of the fifteen objects in the CDF-S sample, Fig. 1 shows that six of them appear to contain enough dust to be detected in the observed sub-mm.  These seven objects, CHa-4, CH8-1, CS2-2, CS2-3, CS2-4 and CS2-6, contain dust in a range A$_{1200}$ = 1.0 -- 4.5 mag.  One of these, CS2-3, which was best-fit by A$_{1200}$ = 4.5 mag, has the brightest f\sub{100} data point (28333 $\mu$Jy) in the whole sample, further proof that our derivation of FIR fluxes is consistent with our model fitting.  Objects CHa-1, CHa-3, CH8-2, CS2-5 and CS2-8 have $\beta$ $<$ -2.5, thus we did not calculate their FIR fluxes.  Objects CHa-2, CS2-1 and CS2-7 do have measurable 60 $\mu$m fluxes (0.2 -- 4.7 $\mu$Jy), but these are likely much too faint to be detectable.  It is of note that CS2-4 is only detectable in the 730 $\mu$m band.

Three of the GRAPES LAEs which we analyzed appear to be bright enough in the FIR to be seen with ALMA, although only in the 730 $\mu$m band.  These objects are 5183, 6139 and 9040.  In Pirzkal et al. (2007), object 5183 has a maximum dust extinction\footnote[2]{This value of dust extinction is the maximum amount of extinction present in this object, because the models were fit to this object using upper limit constraints on the rest-frame optical fluxes.} of A\sub{V} = 0.6 magnitudes (equivalent to A\sub{1200} $\sim$ 2.4), and object 6139 is best fit by 0.4 magnitudes of dust extinction.  This is consistent with our results, which imply that the UV light absorbed by this dust is being re-radiated in the FIR.  While object 9040 was not best-fit by any dust, the authors do acknowledge that its best-fit model was still a poor fit, although they are confident that it is a LAE at its measured redshift.  An important point is that the CDF-S LAEs have a higher detection fraction than the GRAPES LAEs.  This is likely due to differences in the sample selection method.  The CDF-S sample was selected via ground-based narrowband imaging (with space-based broadband photometry).  However, the GRAPES sample was selected via space-based grism spectroscopy over a much smaller area (only the HUDF vs. the whole CDF-S), thus probing much deeper into the faint end of the luminosity function.  As the detectability of an object scales not only with its color, but with its continuum brightness as well, fainter objects will be more difficult to detect.
%-----------------------Figure------1--------------------
\begin{figure}
\psfig{file=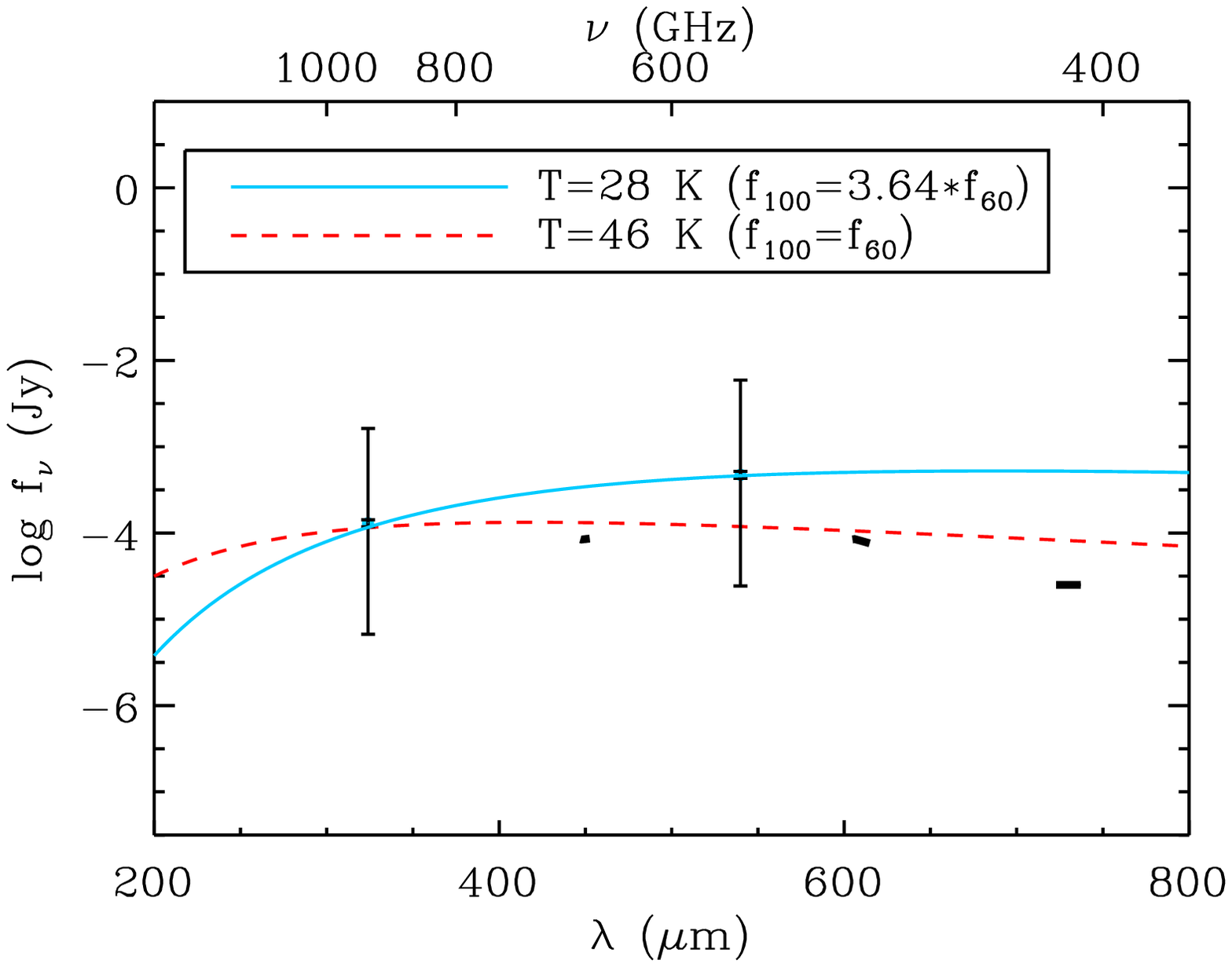,width=3.5in}
\caption{Same plane as Figure 1.  The blue line plots the predicted rest-frame FIR SED for CS2-2.  The literature shows that the temperature of the SED depends on whether the star formation proceeds in compact or widespread environments.  We thus show (in red) our predictions if we assumed a warmer SED, with T = 46K.  The change in temperature does not affect the detection of this object, although it does move the peak of the dust emission towards lower wavelengths.  This can affect our results when an object is only detected in the reddest ALMA band.  We also show the derived errors on our original SED, obtained from the i' and z' photometric errors.  The tiny error bars inside the photometric errors are the errors derived from the scatter in the M97 relation (Meurer et al. 1999).}
\end{figure}
%--------------------------------------------------------

\subsection{Lyman Break Galaxies}
Figure 1 also plots the derived FIR SEDs of the sample of 21 LBGs.  Dust emission from 11 of the LBGs are detectable at 450, 610 $\mu$m and 730 $\mu$m, with two additional LBGs being detectable at only 730 $\mu$m.  In a rough comparison between LBGs and LAEs, 39\% of the 23 LAEs are detectable with ALMA, compared with 62\% for LBGs.  While these numbers may be consistent given the relatively small sample sizes, it could be indicative of a larger difference between the types of objects.  Traditionally, LBGs are thought to harbor more evolved and more massive stellar populations than those in LAEs.  However, from this result, we now see that LBGs may also have a higher dust content than LAEs.  This is confirmed by the average $\beta$ of the two samples, which indicates that LBGs have a redder UV slope with $\left<\beta_{LBG}\right>$ = -1.62, compared with $\left<\beta_{LAE}\right>$ = -2.03.  A similar effect was seen in Vanzella et al. (2006), who found $\left<\beta_{LBG}\right>$ = -1.7 and $\left<\beta_{LAE}\right>$ = -2.1, as well as in Pentericci et al. (2007), who found that LBGs were dustier than LAEs.  This is interesting, given that in F09 we found that our sample of LAEs had much higher dust extinction rates than we would have thought.  Thus, while LAEs appear to have some amount of processed ISM, the ISM of LBGs appears to be much more enriched, resulting in higher dust emission in the FIR.

\subsection{Robustness of Results}
We have chosen a value of f$_{100}$/f$_{60}$ = 3.64 from Bendo et al. (2006), which corresponds to a dust temperature of 28 K.  However, there is a large uncertainty in the FIR color.  Larger values of f$_{100}$/f$_{60}$ imply a cooler ISM (thus a cooler dust temperature), while smaller values imply a warm ISM, more consistent with compact star formation (Dale et al. 2001).  Dale et al. (2001) and Dale and Helou (2002) compute model FIR SEDs assuming a large range in FIR color, from 0.8 $<$ f$_{100}$/f$_{60}$ $<$ 3.3.  Since LAEs are expected to be undergoing significant bouts of star formation, and they are quite compact (e.g., Pirzkal et al. 2007), it is interesting to see how our derived FIR SEDs would change if we assume a warmer value of f$_{100}$/f$_{60}$.  The observations presented in Dale \& Helou (2002) show that the number of objects drops off considerably below f$_{100}$/f$_{60}$ = 1, so we assumed f$_{100}$ = f$_{60}$, and redid our calculations.  Figure 2 shows these results for CS2-2, which was one of the fainter LAES detected in all ALMA bands.  Since warmer FIR SEDs will result in fainter fluxes in the ALMA observable range, this plot can give us an idea of the limits on our predicted fluxes.

We find that although the shape of the SED has changed as expected, this particular object would still be detected.  However, since the peak of the dust emission has moved toward shorter wavelengths, objects which were only detected in the 730 $\mu$m band may be less likely to be detected if they have a warmer SED.  As another illustration of our uncertainties, we have plotted the 1 $\sigma$ errors on our original predicted 60 and 100 $\mu$m fluxes, obtained from the photometric i' and z' errors.  We also plot error bars derived from the uncertainty in the M97 relation, as described in Meurer et al. (1999; 0.04 dex uncertainty in log(F$_{FIR}$/F$_{UV}$)).  These errors are much smaller than the photometric errors.  However, it is worth deeper investigation into the extent which the photometric errors can alter our results.

Using equations 1 -- 6, we can calculate the dust emission from any object given a $z'$ magnitude and a colour ($i' - z'$ for redshifts less than 4.5, and $J - H$ for redshifts greater than 4.5).  Figures 3 and 4 show these results at 730 $\mu$m, with contours showing the flux levels for any colour-magnitude combination, and one contour specifically showing the ALMA 4 hour sensitivity at 730 $\mu$m.  Figure 3 plots all objects in our whole LAE+LBG sample with z $\leq$ 4.53.  The mean redshift of this subsample is 4.33, so the flux curves were derived for this redshift.  Figure 4 plots all objects with z $>$ 4.53, with the flux curves derived for the mean redshift of this subsample, which was 5.26.  

The different coloured circles represent the three different samples (CDF-S LAEs, HUDF LAEs, and HUDF LBGs), with filled circles representing objects we found that would be detectable with ALMA via Figure 1.  All filled circles lie above the ALMA 4 hour sensitivity contour (black line) showing that Figures 1 and 3 \& 4 are mutually consistent, although the three detectable HUDF LAEs do lie less than 1$\sigma$ above the ALMA 4 hour sensitivity contour.  Using these figures, one could then take the colour, magnitude and rough redshift of any object, and determine if the dust emission could be detected by seeing if the points lie above the ALMA sensitivity contour.

Figures 3 \& 4 allow us to examine the robustness of our results, as we have visualized the 1$\sigma$ errors on our objects fluxes.  We can thus see how close our objects are to the ALMA sensitivity limit as a function of their photometric errors.  While 39\% of the LAEs will have their dust emission detectable, only 17\% (4/23) of them lie more than 1$\sigma$ from the ALMA sensitivity contour.  Likewise, for LBGs, only 33\% lie more than 1$\sigma$ from the ALMA 4 hour sensitivity contour.  However, some objects which lie below the ALMA contour are less than 1$\sigma$ away.  Thus, we can define a maximum percentage of objects with detectable dust emission, including not only those above the line, but those that are below but within 1$\sigma$.  We find this to be 61\% for LAEs and 81\% for LBGs.  This allows us to put an error bar on our LAE detection fraction, of 39 $\pm$ 22 \% (likewise 62 $^{+19}_{-29}$ \% for LBGs).
%-----------------------Figure------1--------------------
\begin{figure}
\psfig{file=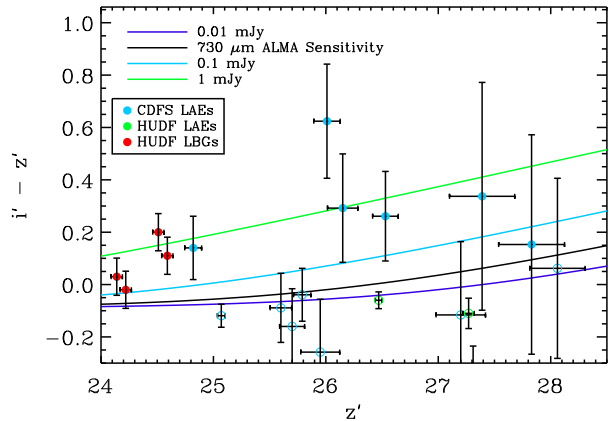,width=3.5in}
\caption{A colour-magnitude diagram, depicting the position of all objects with z $\leq$ 4.53.  Filled circles represent objects which would be detected with ALMA at 730 $\mu$m (and possibly also at 450 and 610 $\mu$m), with open circles representing those objects which would be undetected (based on Figure 1).  Blue circles represent the CDFS LAEs, while red denotes the HUDF LAEs and red the HUDF LBGs.  The contours show the observed fluxes at 730 $\mu$m for a computed grid of magnitudes and colours ($i' - z'$).  The black contour shows the 4 hour ALMA 730 $\mu$m sensitivity limit.  The contours were all computed for z = 4.33, which is the mean redshift of all objects in our sample at z $\leq$ 4.53.  The error bars on the data points are the 1$\sigma$ errors.}
\end{figure}
%--------------------------------------------------------
%-----------------------Figure------1--------------------
\begin{figure}
\psfig{file=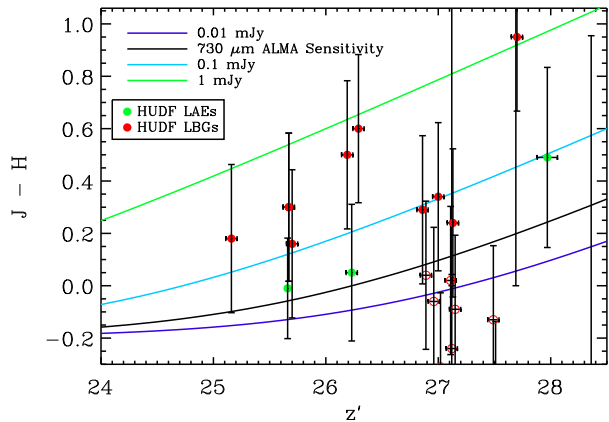,width=3.5in}
\caption{The same as Figure 3, for all objects with z $>$ 4.53, thus the colour is J - H.  The contours were all computed for z = 5.26, which is the mean redshift of all objects in our sample at z $>$ 4.53.}
\end{figure}
%--------------------------------------------------------

While the FIR -- $\beta$ relation was derived at low redshift, both Seibert, Heckman \& Meurer (2002) and Reddy \& Steidel (2004) find that it also applies at high-redshift.  In addition, Teplitz et al. (2006) found that when they compared FUV to Spitzer observations, the results were in general agreement with the FIR -- $\beta$ relation.  Chapman et al. (2005) have investigated this relation using a sample of sub-millimeter selected galaxies at z $\sim$ 2.  The authors conclude that the bolometric luminosity of sub-millimeter galaxies are underestimated by 2 orders of magnitude when derived from their UV properties. This indicates that our derived FIR fluxes may be lower limits, meaning that if anything we could be underestimating the detection fraction.  

However, Reddy et al. (2006) found that the M97 relation may not apply to all star-forming galaxies at z $\sim$ 2.  They find that galaxies with derived ages $<$ 100 Myr exhibit a smaller FIR excess than that predicted by the M97 relation, indicating that these young galaxies have redder UV colors for a given amount of dust extinction that older galaxies.  This is of concern as most of our LAEs have very young derived stellar population ages.   In addition, Baker et al. (2006) find that the M97 relation also over-predicts the FIR flux for the gravitationally lensed LBG cB58.  However, there are differences between these objects and LAEs, as they are at a lower redshift, and they are continuum selected.  We thus cannot know if the same age dependence will apply to LAEs until we obtain a statistical sample of rest-frame FIR detections of LAEs.  In addition, the Reddy et al. figure which illustrates the age discrepancy does not have error estimates on their young object points, thus it is hard to discern how significant this effect is.  Baker et al. do show error bars, and these errors imply that their observations are only off the M97 relation by $\sim$ 1 $\sigma$.  

If young objects do have systematically lower F$_{FIR}$/F$_{UV}$ ratios, it would be worrying, as it could imply that we are over-predicting the FIR fluxes of our LAEs by as much as an order of magnitude.  It is thus prudent to apply our results to any relevant observations to see if our FIR flux predictions are accurate.  Although very few galaxies with \lya~emission have been detected in their rest-frame FIR, Capak et al. (2008) have recently published a study of a spectroscopically confirmed millimeter starburst galaxy, which does exhibit \lya~emission.  They publish the observed SED of this galaxy from X-rays to radio wavelengths.  Using the published optical magnitudes, we can compute $\beta$, and predict the FIR flux.  Doing this, we calculate a predicted 554 (rest 100) $\mu$m flux of 2637$^{+5049}_{-1875}$ $\mu$Jy.  However, the closest point in the observed SED is at 1.1mm, at 4800 $\pm$ 1500 $\mu$Jy.  The flux of our derived SED at 1.1mm is 2054$^{+3932}_{-1460}$ $\mu$Jy, thus they are consistent within 1 $\sigma$.  These observations give evidence that our predictions do match observations.  We show the SED from Capak et al. along with our predictions for this object in Figure 5.  Additionally, we also show the V, i', z' and 3.6 $\mu$m fluxes for CH8-1, which was one of our redder LAEs.  These points have been scaled to match the i' flux of the Capak et al. object.  We can see that the i' - z' color is somewhat redder than the Capak et al. object (0.3 versus 0.15), which indicates that CH8-1 is dustier, explaining its higher expected FIR flux.  However, as it is intrinsically fainter (the scaling factor was $\sim$ 17), it would be much harder to detect with current instrumentation.

Although we have assumed a reasonable SED shape, it is possible that the FIR SED could be much different.  Houck et al. (2004) discuss the low metallicity blue compact dwarf galaxy SBS 0335-052, which has a FIR spectrum much different than typical starburst galaxies, peaking at $\sim$ 28 $\mu$m, indicating much warmer dust, possibly due to the lower metallicity.  High-redshift LAEs may be expected to have a very low metallicity, thus it is worth mentioning that if any of our objects had this type of SED, we could be vastly overestimating their FIR flux.  We have thus investigated the dependence of our results on these derived FIR fluxes.  To do this, we reduced our FIR flux to be 50\% of its derived value, and computed the ALMA detection fraction of both types of objects.  We calculated a detection fraction of 26\% for LAEs and 62\% for LBGs (57\% and 76\% for LAEs and LBGs respectively within 1$\sigma$ of the ALMA contour).  Thus, while we believe that the M97 relation is applicable to our objects, we find that our conclusions are robust to a factor of two variation in the FIR flux. 
%--------------------------------------------------------
\begin{figure}
\psfig{file=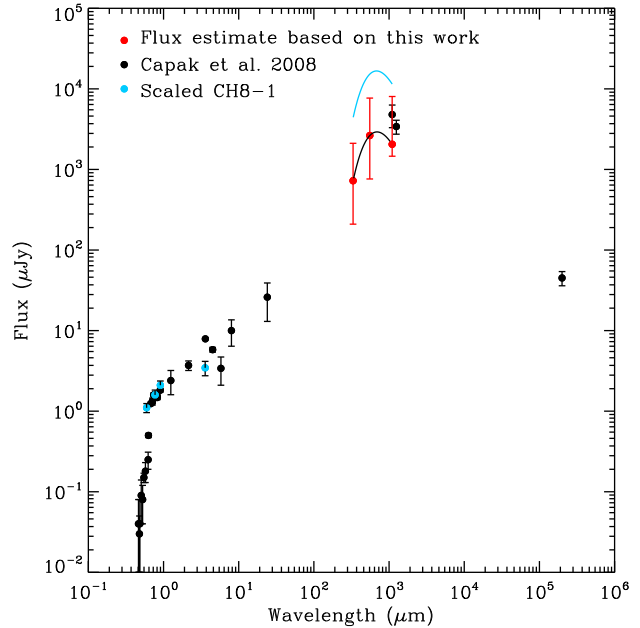,width=3.5in}
\caption{In this figure we plot the SED from Capak  et al. (2008) for a millimeter detected \lya~emitting starburst galaxy, spectroscopically confirmed to lie at z=4.547.  The predicted FIR fluxes match up well with the observed SED, providing further evidence that the M97 relation is applicable at high redshifts.  The blue points denote the V, i', z' and 3.6 $\mu$m fluxes from CH8-1.  These points have been scaled to match the i' flux of the Capak et al. object.  We can see that the i' - z' color is redder than the Capak et al. object, which indicates that CH8-1 is dustier, which we can see as its (scaled) predicted FIR flux is brighter than that of the Capak et al. object.  However, this object is fainter by about a factor of $\sim$ 17, thus it would be much harder to detect with current instrumentation.}
\end{figure}
%--------------------------------------------------------
\section{Conclusions}

Directly observing dust emission from \lya~galaxies will provide undeniable proof that these earliest of galaxies have undergone some enrichment of their ISM.  While these observations are currently difficult, we have been able to predict whether a sample of 23 LAEs would be detected with a future sub-mm observatory such as ALMA based on their optical and NIR fluxes.  Exploiting a relation from Meurer et al. (1997), which relates the far-infrared excess to the UV continuum slope, we have derived the FIR fluxes from these LAEs, which will be primarily due to dust emission.  By comparing these fluxes to the sensitivity of ALMA at 450, 610 and 730 $\mu$m, we have determined that 39\% (9/23) of our sample of LAEs should be detectable based on their dust emission.

We perform the same analysis on a sample of 21 Lyman Break Galaxies, finding that 62\% (13/21) of the LBGs are detectable with ALMA, $\sim$ 60\% higher than the detection fraction of LAEs.  This may be a function of our small sample size, or it may be the result of a true dichotomy in the dust contents of LAEs and LBGs, providing further proof that LBGs are a more evolved class of high-z galaxy.  In order to generalize our results, we computed the FIR fluxes for a wide range of continuum magnitudes and colours, comparing these fluxes to the ALMA 4 hour sensitivity limit.  Thus, knowing the continuum magnitude and colour of any object, one can then predict whether its dust emission will be detectable.  

While stellar population modeling gives some insight on the amount of dust in high-redshift galaxies, future sub-mm instruments will allow us to directly observe the dust emission from these objects, providing a more complete census of dust.  From our analysis of LAEs and LBGs in the HUDF, we calculate a number density of ALMA detectable LAEs and LBGs of 0.26 arcmin\super{-2}, and 1.13 arcmin\super{-2} respectively.  Thus, even in a small survey area, one could expect to detect a number of high-redshift objects with ALMA in a reasonable amount of time.  In our previous analysis of LAEs, the addition of the clumpiness parameter allows strong \lya~flux to an arbitrarily old and dusty galaxy, thus these ALMA observations will help to constrain the amount of dust, and thus the level of clumpiness.

These results will have considerable impact on several areas of galaxy evolution, specifically on the star formation rate (e.g., Giavalisco et al. 1996), which, being derived from UV observations, is critically dependent on accurate measurements of the dust content.  Combining ALMA with UV observations, we will have a complete map of star formation in high-redshift galaxies (i.e. Gao 2008).  Studying the sub-millimeter emission of these objects will also allow calculation of the total dust mass, which is something one cannot calculate from UV-optical data alone.  Along with learning about the total amount of dust in these objects, ALMA will give us the opportunity to investigate the distribution of dust due to its excellent resolution.  At 730 GHz, ALMA can achieve resolutions down to 0\farcs01 pix\super{-1}.  Combining ALMA observations with high resolution {\it HST} data will allow us to investigate dust on local scales at high-redshift for the first time.  Strong sub-millimeter detections combined with strong \lya~emission could indicate that dust enhancement of the \lya~EW is widespread.

\section*{Acknowledgments}
We  thank  the anonymous  referee  for  their  comments which  greatly
improved  the fidelity  of  this  paper.  Support  for  this work  was
provided  in   part  by  NASA  through   grant  numbers  HST-AR-11249,
HST-GO-10240  and  HST-GO-10530   from  the  SPACE  TELESCOPE  SCIENCE
INSTITUTE, which  is operated by  the Association of  Universities for
Research  in Astronomy,  Inc., under  NASA contract  NAS5-26555.  This
work  was  also  supported  by  the  Arizona  State  University  (ASU)
Department  of  Physics  and  the   ASU  School  of  Earth  and  Space
Exploration.


\vspace{-0.1cm}

\begin{thebibliography}{}
\bibitem[\protect\citeauthoryear{Baker et al.}{2006}]{baker} Baker, A. J. et al. 2006, A\&A, 372, L37
\bibitem[\protect\citeauthoryear{Beckwith et al.}{2006}]{hudf} Beckwith, S. V. W. et al. 2006, AJ, 132, 1729
\bibitem[\protect\citeauthoryear{Bendo et al.}{2006}] {bendo} Bendo, G. J. et al 2006, ApJ, 652, 283
\bibitem[\protect\citeauthoryear{Bruzual \& Charlot}{2003}]{bc03} Bruzual, G. \& Charlot, S. 2003, MNRAS, 344, 1000
\bibitem[\protect\citeauthoryear{Calzetti et al.}{1994}]{cal} Calzetti, D., Kinney, A. L. \& Storchi-Bergmann, T. 1994, ApJ, 429, 582
\bibitem[\protect\citeauthoryear{Chapman et al.}{2005}]{chap} Chapman, S. C., Blain, A. W., Smail, I. \& Ivison, R. J. 2005, ApJ, 622, 772
\bibitem[\protect\citeauthoryear{Chary et al.}{2005}]{chary} Chary, R.-R., Stern, D. \& Eisenhardt, P. 2005, ApJ, 635, L5
\bibitem[\protect\citeauthoryear{Capak et al.}{2008}]{capak} Capak, P. et al. 2008, ApJ, 681, L53
\bibitem[\protect\citeauthoryear{Dale et al.}{2001}]{dale} Dale, D. A. et al. 2001, ApJ, 549, 215
\bibitem[\protect\citeauthoryear{Dale \& Helou}{2002}]{dh02} Dale, D. A. \& Helou, G. 2002, ApJ, 576, 159
\bibitem[\protect\citeauthoryear{Dawson et al.}{2007}]{daw07} Dawson, S., Rhoads, J. E., Malhotra, S., Stern, D., Wang, J. X., Dey, A., Spinrad, H. \& Jannuzi, B. T. 2007, ApJ, 671, 1227
\bibitem[\protect\citeauthoryear{Dawson et al.}{2004}]{daw04} Dawson, S., Rhoads, J. E., Malhotra, S., Stern, D., Dey, A., Spinrad, H., Jannuzi, B. T., Wang, J. X. \& Landes, E. 2004, ApJ, 617, 707
\bibitem[\protect\citeauthoryear{Finkelstein et al.}{2007}]{f07} Finkelstein, S. L., Rhoads, J. E., Malhotra, S., Pirzkal, N. \& Wang, J. X. 2007, ApJ, 660, 1023
\bibitem[\protect\citeauthoryear{Finkelstein et al.}{2008a}]{f08a} Finkelstein, S. L., Rhoads, J. E., Malhotra, S., Grogin, N. A. \& Wang, J. X. 2008a, ApJ, 678, 655
\bibitem[\protect\citeauthoryear{Finkelstein et al.}{2009}]{f09} Finkelstein, S. L., Rhoads, J. E., Malhotra, S. \& Grogin, N. A. 2009, ApJ Accepted, astroph/0806.3269
\bibitem[\protect\citeauthoryear{Gao}{2008}]{gao} Gao, Y. 2008, Nature, 452, 417
\bibitem[\protect\citeauthoryear{Giavalisco et al.}{2004}]{gia04} Giavalisco, M., Koratkar, A. \& Calzetti, D. 1996, ApJ, 466, 831
\bibitem[\protect\citeauthoryear{Grazian et al.}{2006}]{music} Grazian, A. et al. 2006, A\&A, 449, 951
\bibitem[\protect\citeauthoryear{Hansen \& Oh}{2006}]{hanoh} Hansen, M. \& Oh, S. P. 2006, MNRAS, 367, 979
\bibitem[\protect\citeauthoryear{Hathi, Malhotra \& Rhoads}{2008}]{nimish} Hathi, N. P., Malhotra, S. \& Rhoads, J. E. 2008, ApJ, 673, 686
\bibitem[\protect\citeauthoryear{Houck et al.}{2004}]{houck} Houck, J. R. et al. 2004, ApJS, 154, 211
\bibitem[\protect\citeauthoryear{Hu et al.}{1998}]{hu98} Hu, E. M., Cowie, L. L. \& McMahon, R. G. 1998, ApJ, 502, L99
\bibitem[\protect\citeauthoryear{Kudritzki et al.}{2000}]{kud00} Kudritzki, R.-P. et al. 2000, ApJ, 536, 19
\bibitem[\protect\citeauthoryear{Lai et al.}{2007}]{lai} Lai, K. et al. 2007, ApJ, 655, 704
\bibitem[\protect\citeauthoryear{Malhotra \& Rhoads}{2002}]{mr02} Malhotra, S. \& Rhoads, J. E. 2002, ApJ, 565, L71
\bibitem[\protect\citeauthoryear{Malhotra et al.}{2003}]{m03} Malhotra, S. Wang, J., Rhoads, J. E., Heckman, T. M. \& Norman, C. A. 2003, ApJ, 585, 25
\bibitem[\protect\citeauthoryear{Malhotra et al.}{2005}]{m05} Malhotra, S. et al. 2005, ApJ, 626, 666
\bibitem[\protect\citeauthoryear{Meurer et al.}{1997}]{meurer97} Meurer, G. R., Heckman, T. M., Lehnert, M. D., Leitherer, C. \& Lowenthal, J. 1997, AJ, 112, 54 {M97}
\bibitem[\protect\citeauthoryear{Meurer et al.}{1999}]{meurer99} Meurer, G. R., Heckman, T. M. \& Calzetti, D. 1999, ApJ, 521, 64
\bibitem[\protect\citeauthoryear{Meier \& Terlevich}{1981}]{mt81} Meier, D. L. \& Terlevich, R. 1981, ApJ, 246, L109
\bibitem[\protect\citeauthoryear{Neufeld}{1991}]{neufeld} Neufeld, D. A. 1991, ApJ, 370, L85
\bibitem[\protect\citeauthoryear{Oke \& Gunn}{1983}]{oke} Oke, J. B. \& Gunn, J. E. 1983, ApJ, 266, 713
\bibitem[\protect\citeauthoryear{Pentericci et al.}{2007}]{pen07} Pentericci, L. et al. 2007, A\&A, 471, 433
\bibitem[\protect\citeauthoryear{Pentericci et al.}{2008}]{pen08} Pentericci, L. et al. 2008, Accepted to A\&A
\bibitem[\protect\citeauthoryear{Pirzkal et al.}{2004}]{grapes} Pirzkal, N. et al. 2004, ApJS, 154, 501
\bibitem[\protect\citeauthoryear{Pirzkal et al.}{2007}]{pirz06} Pirzkal, N. et al. 2007, ApJ, 667, 49 {P07}
\bibitem[\protect\citeauthoryear{Reddy \& Steidel}{2004}]{rs} Reddy, N. A. \& Steidel, C. C. 2004, ApJ, 603, L13
\bibitem[\protect\citeauthoryear{Reddy et al.}{2006}]{reddy} Reddy, N. A. et al. 2006, ApJ, 644, 792
\bibitem[\protect\citeauthoryear{Rhoads et al.}{2000}]{rh00} Rhoads, J. E., Malhotra, S., Dey, A., Stern, D., Spinrad, H. \& Jannuzi, B. T. 2000, ApJ, 545, L85
\bibitem[\protect\citeauthoryear{Rhoads et al.}{2008}]{rh08} Rhoads, J. E. et al. 2008, submitted to ApJ, astroph/0805.1056
\bibitem[\protect\citeauthoryear{Salpeter}{1955}]{salp} Salpeter, E. E. 1955, ApJ, 121, 161
\bibitem[\protect\citeauthoryear{Shimasaku et al.}{2006}]{shima06} Shimasaku, K. et al. 2006, PASJ,58, 313
\bibitem[\protect\citeauthoryear{Seibert, Heckman \& Meurer}{2002}]{shm} Seibert, M., Heckman, T. M. \& Meurer, G. R. 2002, AJ, 124, 46
\bibitem[\protect\citeauthoryear{Spergel et al.}{2007}]{sper} Spergel, D. N. et al. 2007, ApJS, 170, 377
\bibitem[\protect\citeauthoryear{Teplitz et al.}{2006}]{tep} Teplitz, H. I. et al. 2006, AJ, 132, 853
\bibitem[\protect\citeauthoryear{Vanzella et al.}{2006}]{van06} Vanzella, E. et al. 2006, astroph/0612182
\bibitem[\protect\citeauthoryear{Wang et al.}{2004}]{wang04} Wang, J. X., Rhoads, J. E., Malhotra, S., Dawson, S., Stern, D., Dey, A., Heckman, T. M., Norman, C. A. \& Spinrad, H. 2004, ApJ, 608, L21
\bibitem[\protect\citeauthoryear{Wilson}{2007}]{wil} Wilson, T. L. 2007, Astrophys Space Sci, Springer, 10.1007
\bibitem[\protect\citeauthoryear{Wuyts et al}{2008}]{wuyts} Wuyts, S. et al. 2008, astroph/0804.0615
\bibitem[\protect\citeauthoryear{Xu et al}{2007}]{xu} Xu, C. et al. 2007, AJ, 134, 169
\end{thebibliography}
\end{document}